\documentclass[psfig]{aa} 

\usepackage{times}
\usepackage{graphics}
\usepackage{latexsym}
\usepackage{epsfig}

\begin{document}
  \headnote{Letter to the Editor}
  
\title{Dust in 3CR radio galaxies:
  On the FR\,1\,--\,FR\,2 difference\thanks{Based on observations
    with the Infrared Space Observatory ISO, an ESA project with
    instruments funded by ESA Member States (especially the PI
    countries: France, Germany, the Netherlands and the United
    Kingdom) and with the participation of ISAS and NASA.} 
}
\author{S.A.H. M\"uller\inst{1}
  \and M. Haas\inst{1}
  \and R. Siebenmorgen\inst{2}
  \and U. Klaas\inst{3}
  \and K. Meisenheimer\inst{3}
  \and R. Chini\inst{1}
  \and M. Albrecht\inst{4}
}
\offprints{Sven M\"uller (smueller@astro.rub.de)}
\institute{
  Astronomisches Institut, Ruhr-Universit\"at Bochum (AIRUB),
  Universit\"atsstr. 150 / NA7, D-44780 Bochum, Germany
  \and
  European Southern Observatory (ESO),
  Karl-Schwarzschild-Str. 2, D-85748 Garching, Germany
  \and
  Max--Planck--Institut f\"ur Astronomie (MPIA),
  K\"onigstuhl 17, D-69117 Heidelberg, Germany
  \and
  Instituto de Astronom\'ia, Universidad Cat\'olica del Norte,
  Avenida Angamos 0610, Antofagasta, Chile
}
\date{Received ; accepted }
\authorrunning{S.A.H. M\"uller et al.}
\titlerunning{Dust in 3CR radio galaxies}

\abstract{
  We compare three 3CR samples of 11 FR\,1 galaxies, 17 FR\,2 galaxies
  and 18 lobe-dominated quasars contained in the ISO Data Archive. 
  In contrast to the powerful FR\,2 galaxies with edge-brightened
  lobes, the low radio power FR\,1 galaxies in our sample do not
  exhibit any high MIR or FIR dust luminosity, which is typical for a
  buried, intrinsically more luminous AGN.
  This consolidates the fact already inferred from optical studies
  that their AGNs have only a relatively low luminosity.
  Also the FR\,1 galaxies show a high FIR/MIR luminosity ratio,
  compared to quasars, suggesting that their FIR luminosity is
  substantially powered by the interstellar radiation field (ISRF)
  of the giant elliptical hosts.
  Finally, we discuss the \mbox{FR\,1\,--\,FR\,2} morphological
  dichotomy.
  FR\,1 galaxies do not have more interstellar matter (ISM) than
  FR\,2s as traced -- on the large scale -- by the cool FIR emitting
  dust and -- in the nuclear region -- by the warm MIR emitting dust.
  Due to the lack of central gas we suggest that the black holes of
  our FR\,1 galaxies are fed at a lower accretion rate than those of
  the FR\,2 galaxies.

  }

\maketitle


\section{Introduction} 
\label{section_introduction}

  Extended radio galaxies are subdivided into two classes according to
  their radio morphology (Fanaroff \& Riley 1974):
  While the FR\,2 radio galaxies show powerful edge-brightened double
  lobes extending far beyond the host galaxy, the FR\,1 galaxies,
  exhibiting also lower luminosities, do not.
  This FR\,1\,--\,FR\,2 dichotomy occurs around a 408 MHz luminosity
  threshold of about 10$^{\rm 24.5}$ W Hz$^{\rm -1}$, which rises with
  the optical luminosity of the host galaxy (Owen \& Ledlow 1994).
  It is still under debate (e.g. Gopal-Krishna \& Wiita 2001), whether
  the dichotomy is due to {\it intrinsic} properties of the AGN
  launching two different types of jets or due to {\it extrinsic}
  properties with similar jet types but different interaction with the
  host galaxy or the circumgalactic environment. 

  The 3CR Catalogue of radio sources at 178\,MHz (Spinrad et al. 1985)
  contains 298 sources, with about 200 belonging to the FR\,1 and
  FR\,2 classes.
  The brightness of the sources makes the 3CR catalogue a well suited
  database for studying the relationship between these AGN classes.

  Optical studies of the central regions provide good spatial
  resolution, but they may be hampered by dust extinction.
  In fact, HST images of 3CR radio galaxies show filamentary and clumpy
  dust absorption features (e.g. de Koff et al. 2000).
  If the extinction is higher in FR\,2s than in FR\,1s, then observed
  differences of optical quantities like those found by Baum et
  al. (1995) may be simply produced by extinction. 
  Furthermore, often the entire amount of dust is underestimated, when
  derived from optical extinction.
  Therefore, mid- and far-infrared (MIR, FIR) observations of dust
  emission are of great value.
  
  Based on a few sparse IRAS detections and mainly upper limit
  constraints Golombek et al. (1988) already noted that the most
  powerful 3CR radio galaxies appear to be much more FIR luminous than
  the low power ones and that this could be related to the
  FR\,1\,--\,FR\,2 dichotomy.
  From a statistical analysis of coadded IRAS photometry of 3CR and B2
  radio galaxies, Impey \& Gregorini (1993) concluded that the more
  powerful and FR\,2 dominated 3CR sample has warmer FIR dust emission
  than the less powerful and FR\,1 dominated B2 sample.

  With the {\it Infrared Space Observatory} ISO (Kessler et al. 2003)
  about 100 3CR sources of various radio-loud AGN types were
  observed with a detection rate above 50\,\%.
  The full census of the ISO 3CR observations has been published by
  Siebenmorgen et al. (2004) focussing on the ISOCAM MIR data and the
  comparison with dust models, and by Haas et al. (2004) focussing on
  the ISOPHOT MIR-FIR data and the orientation-dependent unification
  of FR\,2 radio galaxies and quasars.
  Corroborating earlier ISO results by Meisenheimer et al. (2001),
  clear evidence was found that the powerful FR\,2 galaxies contain a
  highly dust-enshrouded quasar-like AGN. 

  The ISO archive contains 11 3CR sources of FR\,1 type with giant
  elliptical hosts. 
  Here we consider what their detailed MIR and FIR SEDs contribute to
  the current picture on the relation between FR\,1 and FR\,2
  galaxies.
  We adopt a $\Lambda$ cosmology with H$_0$ = 71 km\,s$^{-1}$\,
  Mpc$^{-1}$, $\Omega_{{\rm m}}$ = 0.27 and $\Omega_{\Lambda}$ =
  0.73. 


\section{Data and Results}
\label{section_data}

  The data are described in Siebenmorgen et al. (2004) and Haas et
  al. (2004).
  They refer to the entire 3CR observations performed by ISO.
  Here we use the randomly collected sample of 11 FR\,1 galaxies, four
  observed in the MIR-FIR by ISOPHOT, and seven observed in the MIR by
  ISOCAM with three of these detected in the FIR by IRAS. 
  The aperture size covers the entire galaxies, being about
  30$\arcsec$ at $\lambda \la 25\,\mu$m, 45$\arcsec$ at 60-100\,$\mu$m
  and 90$\arcsec$ at 120-200\,$\mu$m. 
  In the following we call the 10-40\,$\mu$m range MIR and the
  40-1000\,$\mu$m one FIR.


  \begin{figure}[t]
    \vspace{0.1cm}
    \epsfig{file=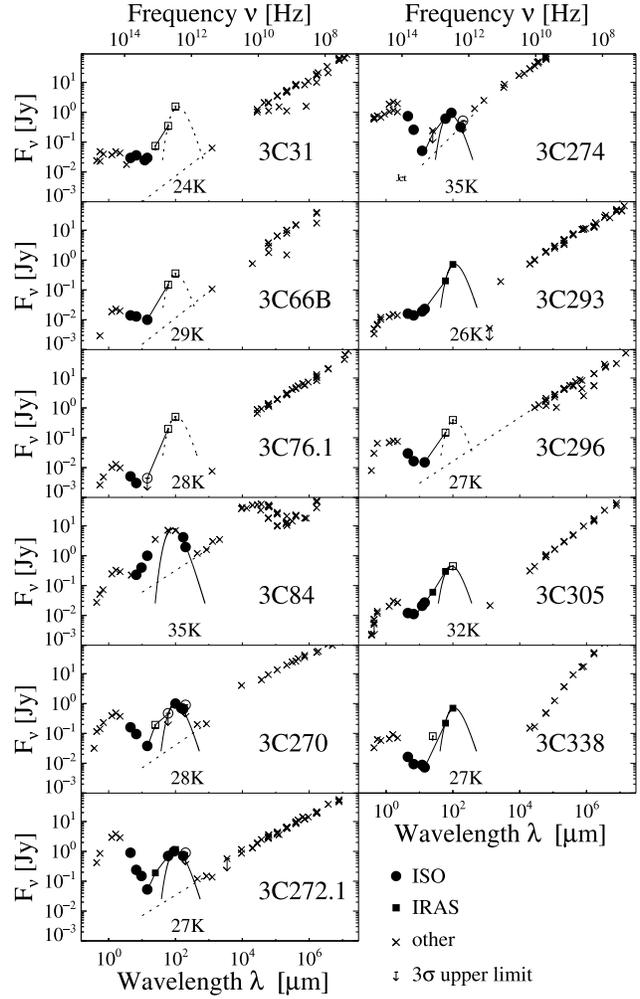, width=8.3cm, clip=true}
    \caption[]{ \label{msxxxx_fig_seds} 
      SEDs of FR\,1 galaxies.
      The measurement errors ($\la$\,30\,\%) are smaller than the
      symbol size. 
      The greybody is interactively fitted to the FIR bump (solid line
      for detections, dotted for upper limits).
      The straight lines connecting the MIR and FIR data points
      indicate the SED envelopes from which the luminosities are
      derived. 
      The straight dashed lines indicate the extrapolation of the
      synchrotron contribution from the cm-mm range to shorter
      wavelengths, if not constrained to low values by the mm data
      points; for 3C\,274 the dashed line shows the contribution of
      the jet as resolved at 12 and 450\,$\mu$m.
      We have reexamined the upper limits at 6.7 and 14.3\,$\mu$m
      (Siebenmorgen et al. 2004), finding clear detections F$_{\rm
      6.7\mu m}$\,=\,11.1\,$\pm$\,2.7\,mJy for 3C\,305, and
      F$_{\rm 14.3\mu m}$\,=\,10.0\,$\pm$\,3.0\,mJy for 3C\,66B.
      3C\,338 has two small companions nearby, which are excluded in
      the ISOCAM photometry but included in the IRAS beam leading to
      a high L$_{\rm FIR}$/L$_{\rm MIR}$ in
      Fig.\,\ref{msxxxx_fig_lmir_to_lfir_vs_lir}.
    }
    \vspace{-0.5cm}
  \end{figure}


  Fig.\,\ref{msxxxx_fig_seds} shows the SEDs of all 11 FR\,1 sources.
  The most remarkable SED features are (1) the stellar optical-NIR
  bump due to the elliptical hosts, (2) the well discernible dust
  emission bump in the MIR-FIR wavelength range, and (3) the steep
  ($\alpha$$>$0.5) synchrotron spectrum in the cm-mm wavelength
  range. 
  The SEDs show two minima, one around 1\,mm, where the synchrotron
  radiation falls below the FIR dust emission, and one at
  $\lambda$$\approx$10\,$\mu$m, where the Rayleigh-Jeans tail of the
  stellar bump meets the MIR-FIR emission bump.
  The $\lambda$$^{\rm -2}$ modified greybodies refer to the cool dust
  component which peaks in the FIR around 100\,$\mu$m. 
  Their temperatures are used to derive the dust mass.
  Table\,\ref{msxxxx_tab_luminosities} lists the luminosities and
  other values derived in the same manner as described by Haas et
  al. (2004).

  For comparison we use the 3CR samples of 17 FR\,2 galaxies and 18
  quasars (including broad line radio galaxies) with clear thermal
  dust bumps (Fig.\,1 and Tab.\,2 in Haas et al. 2004).

  \section{Discussion}
  \label{section_discussion}
  The strength of the AGN in FR\,1s is believed to be low.
  HST images show only marginal evidence for thick obscuring dust tori
  in FR\,1s, but optical observations may have missed the obscuring
  material.
  Therefore, we perform a check in the IR.


  \begin{table}[h!]
    \vspace{-0.5cm}
    \caption[] { Parameters of the FR\,1 galaxies, derived in the same
      way as by Haas et al. (2004).
      L$_{\rm eml}$=L(H$_\alpha$+[NII]) is adopted from Zirbel \& Baum
      (1995).
      \label{msxxxx_tab_luminosities} 
    } 
    \vspace{-0.2cm}
    \scriptsize
    \begin{tabular}{@{\hspace{0.5mm}}r@{\hspace{1.1mm}}r@{\hspace{2.1mm}}r@{\hspace{2.1mm}}r@{\hspace{2.1mm}}r@{\hspace{2.1mm}}r@{\hspace{2.1mm}}r@{\hspace{1.1mm}}r@{\hspace{1.1mm}}r@{\hspace{1.1mm}}}
      3CR & D$_{\rm L}$ & L$_{\rm MIR}$  & L$_{\rm FIR}$& P$_{\rm 15 \mu m}$ & P$_{\rm 178\,MHz}$ & T$_{\rm dust}$ & M$_{\rm dust}$ & L$_{\rm eml}$\\ 
      &   { Mpc}   &
      { log\,[L$_{\odot}$]}&
      { log\,[L$_{\odot}$]}&
      { log\,[W]}&
      { log\,[W]}&
      { [K]}&
      { log\,[M$_{\odot}$] } & { log\,[W]}\\
\hline
 31.0 & 70&$<$9.23& $<$9.84&$<$35.55&33.25&$\sim$24&$\sim$7.00&  ...\\
 66\,B& 91&$<$8.85& $<$9.47&$<$35.10&33.64&$\sim$29&$\sim$6.23&  ...\\
 76.1 &136&$<$9.21& $<$9.97&$<$35.35&33.70&$\sim$28&$\sim$6.82&  ...\\
 84.0 & 74&   0.87&   10.80&   37.11&33.85&      35&      7.01&34.91\\
270.0 & 31&   8.84&    9.05&   35.10&33.03&      28&      5.81&32.76\\
272.1 & 12&   8.13&    8.33&   34.46&31.90&      27&      5.13&32.37\\
274.0 & 16&   8.43&    8.50&   34.72&32.56&      35&      4.75&33.47\\
293.0 &197&   9.98&   10.43&   36.32&34.04&      26&      7.41&34.55\\
296.0 &104&$<$9.24& $<$9.64&$<$35.57&33.45&$\sim$27&$\sim$6.51&  ...\\
305.0 &179&   9.97&   10.23&   36.29&34.04&      32&      6.66&33.12\\
338.0 &123&   9.29&$<$10.02&   35.52&34.21&      27&      6.95&33.77\\
\hline
    \end{tabular}
    \vspace{-0.2cm}
  \end{table}
   

  The FR\,1s exhibit rather weak MIR and FIR luminosities below
  10$^{\rm 10}$ L$_{\odot}$ (Tab.\,\ref{msxxxx_tab_luminosities}).
  If the FR\,1 sources had powerful AGN, one should see their
  signatures either directly via the broad line regions or via the
  reemission from  absorbing dust.
  But they neither exhibit a MIR luminosity which would be typical
  for a buried powerful AGN (L$_{\rm MIR}$$\ga$10$^{\rm
  11}$L$_{\odot}$), nor do they show any prominent optical broad line
  region signatures.
  This confirms the known conclusion (Baum et al. 1995, Whysong \&
  Antonucci 2004) that the low luminosity 3CR FR\,1 galaxies contain
  only a low luminosity AGN.

  The FIR luminosity of the detected FR\,1s is probably dominated by
  dust emission, since the extrapolation of the extended
  synchrotron spectrum reaches at most 10\% of the FIR fluxes.
  Even in the case of a flatter nuclear synchrotron spectrum its
  contribution to the FIR flux peak might be below 50\,\%.
  However, the MIR luminosity of FR\,1s may be due to three
  components, namely the stellar hosts, synchrotron radiation and dust
  emission. 
  The latter one may be powered by the AGN, starbursts and the
  interstellar radiation field (ISRF).
  In the following we consider, how  these contributions can be
  discriminated by means of integral photometry.

  At wavelengths $\lambda$\,$>$\,10\,$\mu$m the stellar blackbody
  steeply declines below the observed MIR fluxes
  (Fig.\,\ref{msxxxx_fig_seds}) contributing at most a few percent to
  the entire MIR luminosities. 
  For the two sources 3C\,270 and 3C\,274, which are well resolved by
  ISOCAM, the spatial extent (FWHM) of the 14.3\,$\mu$m and
  12.0\,$\mu$m emission is by a factor of two smaller than that at
  4.5\,$\mu$m. 
  These two facts argue against a significant stellar contribution to
  the MIR luminosity.

  The synchrotron contribution, extrapolated from the cm-mm range
  (Fig.\,\ref{msxxxx_fig_seds}), lies in most cases below 30\% of the
  MIR fluxes.
  An exception is the extreme case of 3C\,274 (M\,87) with a prominent
  extended optical jet.
  The resolved contribution of the jet reaches about 30\% and 70\% of
  the flux at 12\,$\mu$m and 450\,$\mu$m, respectively (Fig. B1 in
  Siebenmorgen et al. 2004, Sect. 3.2 in Haas et al. 2004).
  In their case study of M\,87 Whysong \& Antonucci (2004) find an
  unresolved nucleus with a 11.7\,$\mu$m flux density of 13\,mJy
  inside a synthetic 0$\farcs$6 aperture.
  This is consistent with our result of 35 mJy at 12\,$\mu$m over a
  larger area ($\sim$20$\arcsec$), excluding the resolved jet flux (16
  mJy).
  The remaining 12\,$\mu$m flux of 22 mJy comes from a larger non-jet
  area, hence has no synchrotron origin,
  but rather originates from a stellar host or dust emission.
      
  The case of M\,87 demonstrates that our MIR luminosities may
  overestimate the MIR {\it dust} luminosity in general.
  For the other sources of our sample no MIR information of high
  spatial resolution is available in the literature.
  From our integral photometry we cannot properly disentangle the
  synchrotron and dust contributions to the MIR luminosity.
  We continue the discussion under the assumption that the major
  fraction of the MIR luminosity of most of our sources is due to dust
  emission.
  

  \begin{figure}
    \vspace{0.1cm}
    \epsfig{file=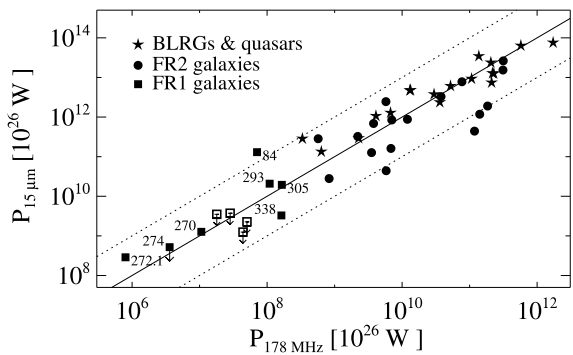, width=7.5cm, clip=true}
    \vspace{-0.1cm}
    \caption[]{ \label{msxxxx_fig_lmir_vs_radio} 
      MIR 15 $\mu$m versus 178\,MHz radio power.
      Open symbols with arrows denote 3$\sigma$ upper limits.
      The lines indicate a ratio of 100 (solid line) and a factor of
      ten around this ratio (dotted lines).
    }
    \vspace{-0.4cm}
  \end{figure}


  Since the FR\,1s show little evidence for luminous dust tori and
  their AGNs are weak, we check whether their MIR emission
  -- if due to dust -- is powered by the AGNs or by the presumably
  strong interstellar radiation field (ISRF) of the giant hosts.
  Starbursts might play only a minor role, as shown further below.
  Therefore we compare the MIR luminosity with two independent AGN
  luminosity measures:

  1) The ratio of 15\,$\mu$m to 178\,MHz power shows the same
  dependency for FR\,1s as for FR\,2s and quasars
  (Fig.\,\ref{msxxxx_fig_lmir_vs_radio}).
  In the FR\,2s and the quasars the MIR power is dominated by AGN
  heated dust emission (Haas et al. 2004).   
  Even, if for the FR\,1s only 30-50\% of the MIR power were due to
  dust, the ratio P$_{\rm 15 \mu m}$\,/\,P$_{\rm 178\,MHz}$ would still
  lie within the dispersion of the distribution of a factor of ten.
  The similarity of this ratio, independent of radio power and source
  type, would be difficult to understand, if in the FR\,1s the MIR dust
  emission were heated only by the ISRF without any significant AGN
  contribution. 
  
  2) We consider the luminosity of the {\it narrow} H$_\alpha$+[NII]
  emission lines (from Zirbel \& Baum 1995) which refers to the
  central 2.5 kpc of the sources.
  The emission line luminosity of FR\,1s, normalized by the absolute V
  magnitude of the host, lies between the high range of FR\,2s and the
  low one of radio-quiet elliptical galaxies (Fig.\,2 in Baum et
  al. 1995). 
  This argues against a pure host galaxy origin of the emission lines
  and favours a significant AGN contribution to the luminosity in
  FR\,2s and also in FR\,1s.  
  Now considering our data, the ratio of emission line luminosity and
  MIR 15\,$\mu$m power is similar for all source types (range
  $10^{-1}$ to $10^{-3}$) and independent of radio power.
  This would be difficult to understand if P$_{\rm 15 \mu m}$ were
  purely due to ISRF heated dust.
  If the emission line luminosity were significantly driven by the AGN,
  then P$_{\rm 15 \mu m}$ should include power from the AGN heated
  dust, too.
  
  These two results are consistent with the interpretation that also
  in the FR\,1s the MIR dust emission can be powered by the AGN, but a
  considerable contribution from the host's interstellar radiation
  field cannot be excluded.


  \begin{figure}
    \vspace{0.1cm}
    \epsfig{file=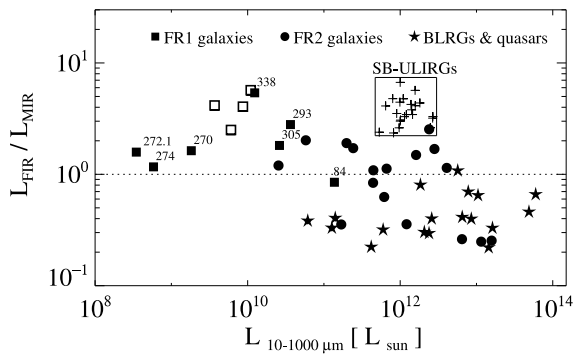, width=7.5cm, clip=true}
    \vspace{-0.15cm}
    \caption[]{ \label{msxxxx_fig_lmir_to_lfir_vs_lir}
      Ratio of FIR to MIR versus IR luminosity.
      Open squares are upper MIR and FIR flux limits.
      The rectangular box contains the starburst ULIRGs (+) (Klaas et
      al. 2001). 
      The dotted line indicates the L$_{\rm FIR}$/L$_{\rm MIR}$
      transition border between FR\,1 galaxies and quasars.
    }
    \vspace{-0.4cm}
  \end{figure}


  As mentioned above, the FIR luminosity of FR\,1s is probably
  dominated by dust emission. 
  Relative to both the MIR and the radio 178\,MHz luminosities,
  the FIR luminosity of the FR\,1 galaxies is higher than that of
  the quasars and many FR\,2 galaxies
  (Fig.\,\ref{msxxxx_fig_lmir_to_lfir_vs_lir}).
  This suggests that in the FR\,1s the FIR luminosity is substantially
  provided by another heating source than by a weak AGN, e.g. at a
  first glance by starbursts.
  But the temperature T of the dust component associated with the FIR
  flux peak lies between 25 K and  35 K for the FR\,1s
  (Fig.\,\ref{msxxxx_fig_seds}).
  This is exactly the range found for radio-quiet elliptical galaxies
  (e.g. Bregman et al. 1998), and lower than for FR\,2s and luminous
  starburst galaxies (T\,$>$\,35\,K).
  Also, the ratio of total optical luminosity, including that from
  the host galaxy, to the FIR luminosity is about 50-500 times
  higher for FR\,1s than for FR\,2s and dusty starburst galaxies.
  Therefore, we conclude that in FR\,1s the bulk of the FIR emission
  is mainly powered by the ISRF of the giant host galaxy.
  If the dust is distributed quite homogeneously, then the ISRF may
  heat it to higher temperatures than those around T = 15-20 K found
  in spiral galaxies, where the bulk of the dust is typically
  concentrated in dense clumps shielding against the ISRF.
  The AGN or starbursts, if any, play only a minor role for the FIR
  luminosity of FR\,1s.
  This is consistent with the results by Heckman et al. (1986) and de
  Koff et al. (2000) that many FR\,2 hosts show morphological
  distortions and signs of interaction which trigger starbursts and
  nuclear activity, while the FR\,1 hosts are typically regular
  ellipticals. 
  
  Finally, in the case of rather homogeneously distributed dust as
  indicated for the FR\,1s one would expect that there is not much
  cold dust, which would radiate at submillimetre wavelengths and
  could have escaped our detection.
  
  In order to explain why the FR\,1s do not show the powerful extended
  radio lobes like FR\,2s, deceleration of the FR\,1 jets due to
  entrainment of the ISM or stellar wind material has been proposed
  (de Young 1993, Bicknell 1994, 1995, Bowman et al. 1996).
  Also the existence of sources with hybrid morphology -- FR\,1 on one
  side, FR\,2 on the other side of the core -- argues against {\it
  intrinsic fundamental physical} differences in the central engine,
  such as black hole spin (slow for FR\,1, fast for FR\,2,) or jet
  plasma composition (e$^{-}$p$^{+}$ for FR\,1, e$^{-}$e$^{+}$ for
  FR\,2), and favours an {\it extrinsic} origin for the
  FR\,1\,--\,FR\,2 dichotomy (Gopal-Krishna \& Wiita 2000).
  In addition to entrainment, the pressure of the external hot X-ray
  emitting gas turns out to play a role in the confinement of the jets
  (e.g. Laing \& Bridle 2002).
  
  Since the edge-darkened FR\,1 lobes are contained within the host
  galaxy, the relevant jet deceleration must take place therein and
  not far outside in the circumgalactic medium.
  But what is the main source for entrainment in FR\,1s?

  With dust masses between 10$^{\rm 5}$ and 10$^{\rm 7}$\,M$_{\odot}$
  the FR\,1s (Tab.\,\ref{msxxxx_tab_luminosities}) do not show
  evidence for more {\it interstellar} matter (ISM) than the FR\,2s
  having dust masses between 10$^{\rm 6}$ and 10$^{\rm
  9}$\,M$_{\odot}$ (Tab.\,2 in Haas et al. 2004).
  If the FR\,1\,--\,FR\,2 dichotomy were {\it purely} due to
  deceleration of the jets {\it by ISM entrainment}, then one would
  expect the FR\,1s to be less affected, contrary to what is
  observed.
  As one possible way out, the ISM in FR\,2s has to be concentrated in
  a special geometry, e.g. avoiding a bi-cone, so that the jets are
  not so much affected.
  Such an explanation, however, is questionable, since on HST images
  FR\,2s show more disturbed dust structures (de Koff et al. 2000),
  while FR\,1s appear more regular.
  Thus, that kind of presumably cool ISM, which we can infer via the
  dust emission, appears not to play a role for the jet entrainment.

  Current models have considered stellar wind material to be
  entrained.
  Such material, even if dusty, may be too marginal to show up in
  our SEDs, which are dominated by the entire dust content.
  In addition to stellar wind material in the giant elliptical hosts
  of FR\,1 galaxies the {\it stars} themselves crossing the jet stream
  could contribute considerably to the material entrained.
  As a rough estimate, assuming a jet opening angle of 1$\degr$ and
  that the 10$^{\rm 12}$ stars of a giant elliptical galaxy are evenly
  distributed over the sphere, about 10$^{\rm 7-8}$ stars lie on the
  jet trajectory, providing plenty of material, of the order of
  10$^{\rm 7-8}$ M$_{\odot}$, for entrainment.
  Then, also the Owen-Ledlow effect can naturally be understood:
  More luminous hosts contain more stars and have a larger velocity
  dispersion; therefore, the chance that stars cross the jet stream and
  decelerate the jet increases with host luminosity.
  So far, however, it is not yet clear, what happens with a star
  approaching the jet stream and whether a star would survive such an
  event (e.g. Bednarek \& Protheroe 1997).

  Furthermore, if the dust traces the amount of ISM and if the warm
  MIR emitting dust indicates also whether there is much ISM in the
  central region, then, with regard to the more dusty FR\,2s, the
  FR\,1s suffer a lack of ISM fuel for the immediate accretion
  region.
  This suggests that the black holes of the FR\,1s are fed at a lower
  accretion rate than those of the FR\,2s. A similar conclusion was
  reached by Baum et al. (1995) from the low emission-line luminosity
  of FR\,1s. 

  Our IR data of 3CR sources suggest that a combination of central as
  well as extended differences should be considered for explaining the
  FR\,1\,--\,FR\,2 dichotomy.
  Furthermore, evolutionary effects could play a role, in the sense
  that the FR\,1 galaxies could be preferentially old AGN with
  starving black holes.
  Since none of the FR\,1 galaxies shows dust properties (like masses
  and temperatures) comparable to those found for the FR\,2s and
  quasars, the results from our small randomly selected samples may be
  valid even more generally.
  Clearly, the phenomena leading to the FR\,1\,--\,FR\,2 dichotomy are
  highly complex, and further data are required, before definite
  conclusions about these suggestions can be drawn.


  \acknowledgements 
  We thank the referee P. Wiita for his constructive suggestions.
  This work was supported by the Nordrhein-Westf\"alische Akademie der
  Wissenschaften.

\end{document}